\documentclass[11pt]{revtex4}
\usepackage{amssymb,epsf}
\usepackage{epsfig}
\usepackage{graphicx, latexsym, amssymb, amscd, psfrag}
\usepackage{latexsym}
\begin{document}

\title{Families of stable and metastable solitons in coupled system of scalar fields}
\author{ Nematollah Riazi$^{1,2}$\footnote{email: riazi@physics.susc.ac.ir} and Marzieh Peyravi$^{1}$}
\affiliation{$1$. Physics Department and Biruni Observatory,
Shiraz University, Shiraz 71454, Iran, \\ and \\ $2$. Physics
Department, Shahid Beheshti University, Tehran 19839, Iran.}

\begin{abstract}
In this paper, we obtain stable and metastable soliton solutions
of a coupled system of two real scalar fields with five five
discrete points of vacua. These solutions have definite
topological charges and rest energies and show classical dynamical
stability. From a quantum point of view, however, the V-type
solutions are expected to be unstable and decay to D-type
solutions. The induced decay of a $V$-type soliton into two
$D$-type ones is calculated numerically, and shown to be chiral,
in the sense that the decay products do not respect left-right
symmetry.
\end{abstract}

\maketitle
\section{Introduction\label{intro}}

Relativistic solitons, including those of the well-known
Sine-Gordon (SG) equation, exhibit remarkable similarities with
classical particles. They exert short range forces on each other
and collide, without losing their identities \cite{0,1,2,3,4}.
They are localized and do not disperse while propagating in the
medium. Because of their field nature, they do tunnel a barrier in
certain cases, although this tunnelling is different from the
well-known quantum version \cite{1,5}.Topological solitons are
stable, due to the boundary conditions at spatial infinity. Their
existence, therefore, is essentially dependent on the presence of
degenerate vacua \cite{1,6}.

Topology provides an elegant way of classifying solitons in
various sectors according to the mappings between the degenerate
vacua of the field and the points at spatial infinity. For the
Sine-Gordon system in $1+1$ dimensions, these mappings are between
$\phi=2n\pi$, $n\in\mathbb{Z}$ and $x=\pm\infty$, which correspond
to kinks and antikinks of the SG system. More complicated mappings
occur in solitons in higher dimensions \cite{2}. Coupled systems
of scalar fields with soliton solutions have found interesting
applications in double-strand, long molecules like the DNA
molecules \cite{dna}-\cite{pey} and bi-dimensional QCD \cite{qcd}.

Bazeia et al. \cite{ba} considered a system of two coupled real
scalar fields with a particular self-interaction potential such
that the static solutions are derivable from first order coupled
differential equations. Riazi et al. \cite{riman} employed the
same method to investigate the stability of the single-soliton
solutions of a particular system of this type. Inspired by the
coupled system introduced in \cite{1}, we propose a new coupled
system of two real scalar fields which shows interesting types of
solitons with well-defined topological charges and rest energies
(masses).

In this paper, we focus on a system with the self-interaction
potential
\begin{equation}\label{aa}
V(\phi,\psi)=\phi^2(\psi^2-\psi_{0}^2)^2+\psi^2(\phi^2-\phi_{0}^2)^2,
\end{equation}
in which $\phi$ and $\psi$ are real scalar fields, and $\phi_{0}$
and $\psi_{0}$ are constants. This potential is plotted in Fig.
\ref{1} for $\phi_{0}=1$ and $\psi_{0}=2$. Following the common
terminology in field theory, this potential has absolute
degenerate minima at $\phi=\psi=0$, $\phi = \pm\phi_{0}$ and
$\psi=\pm\psi_{0}$ known as the true vacua. In our proposed
system, a spectrum of solitons with different rest energies exists
which are stable or meta-stable, depending on their energies and
boundary conditions.

The structure of this paper is as follows: in Section \ref{sec2}
we review some basic properties of the proposed system. In Section
\ref{sec3}, some exact solutions together with the corresponding
charges and energies are derived. The necessary nomenclature and
general behavior of the solutions due to the boundary conditions
are also introduced in this section. Numerical solutions
corresponding to different boundary conditions are presented, and
properties of these solutions like their charges, masses, and
stability status are addressed in this section. In order to
investigate the stability of the numerical solutions, their
evolution is worked out numerically. The classical dynamical
stability of the solutions is investigated further in \ref{sec4}.
Our conclusions and a summary of the results are given in the last
section \ref{sec5}.

\begin{figure}[h]
\epsfxsize=8cm\centerline{\hspace{8cm}\epsfbox{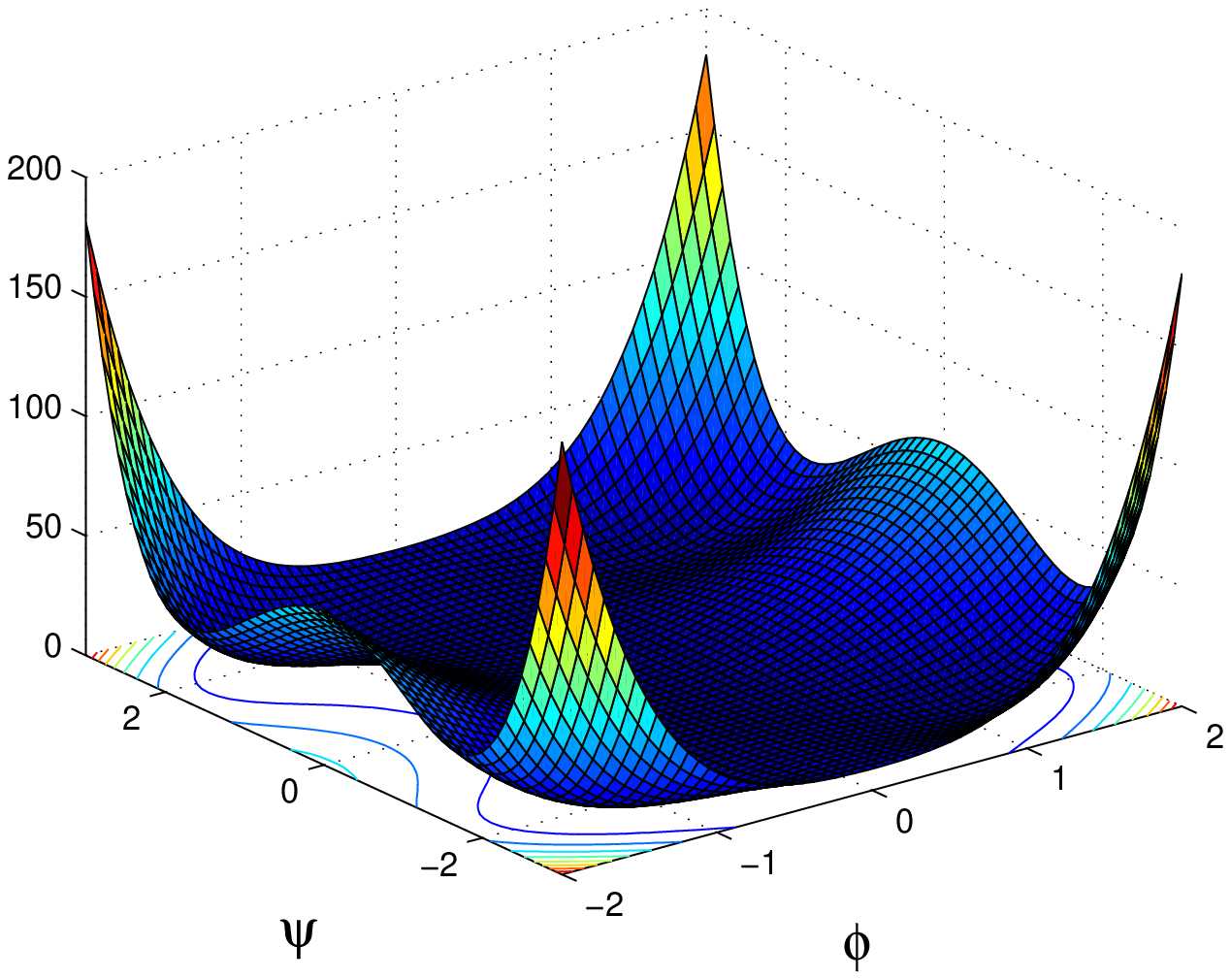}\epsfxsize=10cm\centerline{\hspace{-7cm}\epsfbox{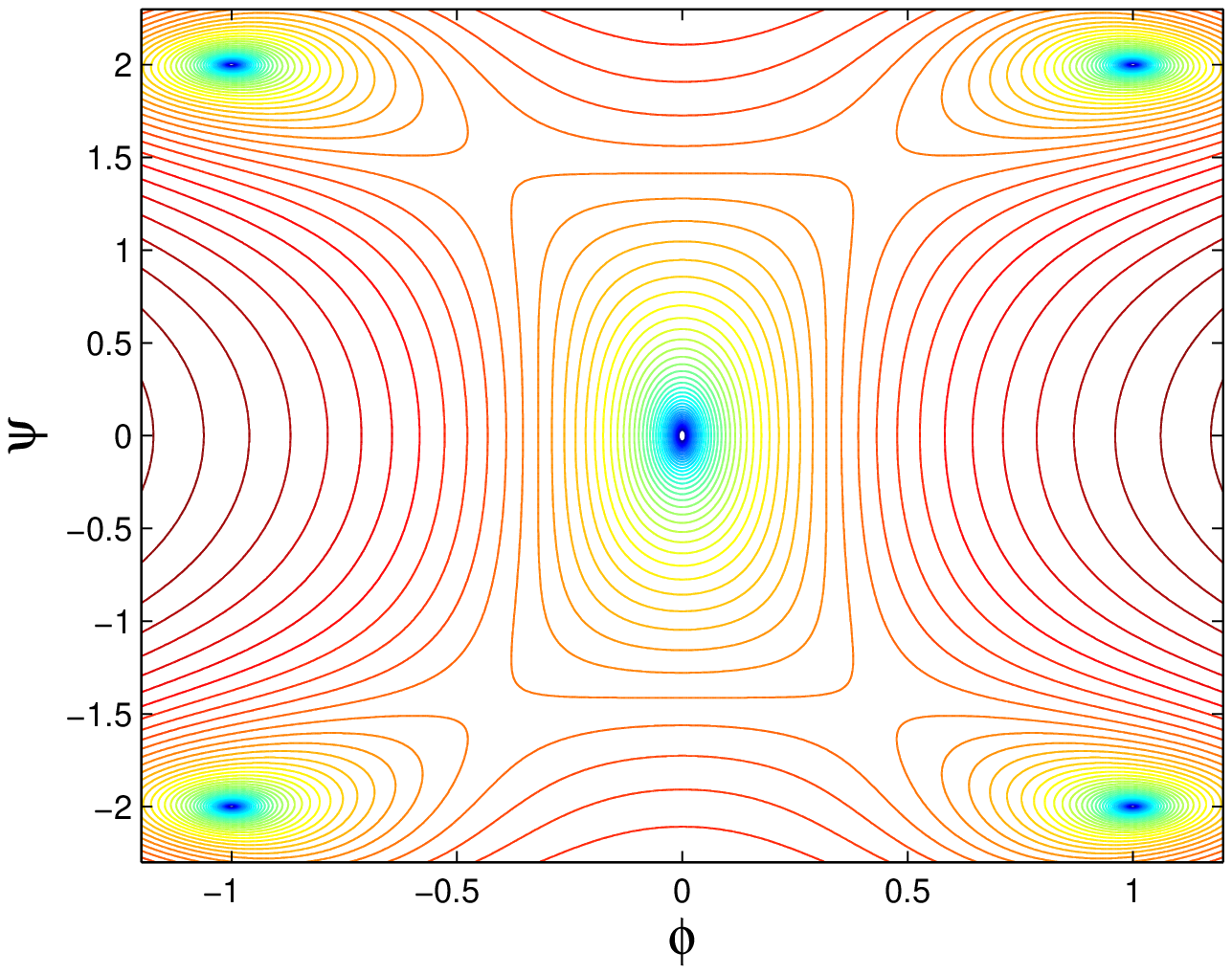}}}
\caption{Left: The self-interaction potential (1) is shown as a
height diagram over the ($\phi$,$\psi$) plane.
$(\phi_0,\psi_0)=(1,2)$ is assumed throughout this paper. Right:
The self-interaction potential as a contour map over the
$(\phi,\psi)$ plane. }
\end{figure}

%\begin{figure}[h]
%\epsfxsize=10cm \centerline{\epsfbox{potential.eps}} \caption{The
%self-interaction potential (1) is shown as a height diagram over
%the ($\phi$,$\psi$) plane. $(\phi_0,\psi_0)=(1,2)$ is assumed
%throughout this paper.\label{1}}
%\end{figure}
%
%
%\begin{figure}[h]
%\epsfxsize=15cm \centerline{\epsfbox{contour.eps}} \caption{The
%self-interaction potential as a contour map over the $(\phi,\psi)$
%plane.\label{cont}}
%\end{figure}
%
%
\section{Dynamical Equations, Conserved Currents, and Topological Charges}\label{sec2}

Within a relativistic formulation, the Lagrangian density of the
system is given by:
\begin{equation}\label{a}
{\cal L}=
\frac{1}{2}\partial^{\mu}\phi\partial_{\mu}\phi+\frac{1}{2}\partial^{\mu}\psi\partial_{\mu}\psi
-[\phi^2(\psi^2-\psi_{0}^2)^2+\psi^2(\phi^2-\phi_{0}^2)^2].
\end{equation}
From this Lagrangian density, we obtain the following equations
for $\phi$ and $\psi$, respectively:
\begin{equation}\label{b}
\Box\phi=-2\phi(\psi^2-\psi_{0}^2)^2-4\phi\psi^2(\phi^2-\phi_{0}^2);
\end{equation}
and
\begin{equation}\label{c}
\Box\psi=-2\psi(\phi^2-\phi_{0}^2)^2-4\psi\phi^2(\psi^2-\psi_{0}^2).
\end{equation}
Since the lagrangian density Eq.(\ref{a}) is Lorentz invariant,
the corresponding energy-momentum tensor\cite{7,8} is:
\begin{equation}\label{d}
T_{\mu\nu} =
\partial_{\mu}\phi\partial_{\nu}\phi+\partial_{\mu}\psi\partial_{\nu}\psi
- g_{\mu\nu}{\cal L};
\end{equation}
which satisfies the conservation law
\begin{equation}\label{e}
\partial_{\mu}T^{\mu\nu} =0.
\end{equation}
In Equation (\ref{d}),  $g_{\mu\nu}=diag(1,-1)$ is the metric of
the $(1+1)$ dimensional Minkowski spacetime. The Hamiltonian
(energy) density is obtained from Eq.(\ref{d}) according to
\begin{equation}\label{f}
\mathcal{H}=T^{00}=\frac{1}{2}\left(\frac{\partial \phi}{\partial
t}\right)^2+\frac{1}{2}\left(\frac{\partial \psi}{\partial
t}\right)^2+\frac{1}{2}\left(\frac{\partial \phi}{\partial
x}\right)^2+\frac{1}{2}\left(\frac{\partial \psi}{\partial
x}\right)^2+V(\phi,\psi).
\end{equation}
It can be shown that the following topological currents can be
defined, which are conserved independently, and lead to quantized
charges:

\begin{eqnarray}\label{g}
J^{\mu}_{H}=\frac{1}{2\phi_0}\epsilon^{\mu\nu}\partial_{\nu}\phi,\nonumber\\
J^{\mu}_{V}=\frac{1}{2\psi_0}\epsilon^{\mu\nu}\partial_{\nu}\psi .
\end{eqnarray}
The subscripts H, V and D (to be used later), denote
``horizontal'', ``vertical'' and ``diagonal'' which will be
explained later. The currents $J^{\mu}_{H,V}$ are conserved,
independent of each other:
\begin{eqnarray}\label{h}
\partial_{\mu}J^{\mu}_{H}=0,\nonumber\\
\partial_{\mu}J^{\mu}_{V}=0.
\end{eqnarray}
It is obvious that the total horizontal and vertical charges are
conserved separately in any local dynamical evolution of the
system. The corresponding topological charges are given by:
\begin{eqnarray}\label{i}
Q_{H}=\int_{-\infty}^{+\infty}J_{H}^{0}dx=\frac{1}{2}[\phi(+\infty)-\phi(-\infty)],\nonumber \\
Q_{V}=\int_{-\infty}^{+\infty}J_{V}^{0}dx=\frac{1}{4}[\psi(+\infty)-\psi(-\infty)].
\end{eqnarray}

\begin{figure}[h]
\epsfxsize=10cm \centerline{\epsfbox{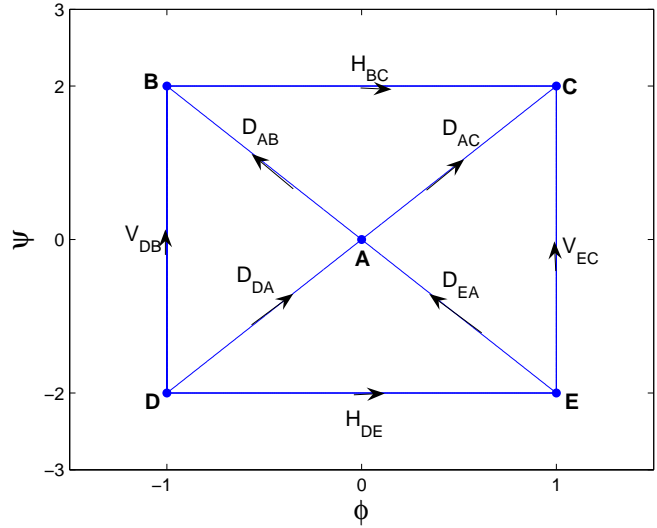}}
\caption{Nomenclature of horizontal (H), vertical (V) and diagonal
(D) solutions according to the boundary conditions on the $(\phi ,
\psi )$ plane.\label{2}}
\end{figure}

\section{Single-soliton solutions}\label{sec3}
Static solutions which correspond to transitions between
neighboring vacua are symbolically shown in Fig. \ref{2}.
Accordingly, we call the static solutions H (horizontal), V
(vertical) and D (diagonal) types, depending on the beginning and
end points of the solution on the $(\phi,\psi)$ plane. We have
called them ``horizontal'', ``vertical'' and ``Diagonal'' simply
because of their orientation in the ($\phi$ ,$\psi$) plane. If we
apply -in turn- the constraints $\psi=\psi_0$, $\phi=\phi_0$, and
$\psi=\frac{\psi_0}{\phi_0}\phi$ in the Lagrangian density
(\ref{a}), the resulting one degree of freedom equations will
reduce to the well-known $\phi^4$ and $\phi^6$ systems which
possess the following exact single-soliton solutions\cite{10,11},
summarized in Table 1. Of course, these are not the exact
solutions of the coupled equations (\ref{b}) and (\ref{c}).

\begin{center}
Table 1. Exact static solutions and the corresponding horizontal
and vertical charges.
\begin{tabular}{|c|c|c|c|}
  \hline
  \hline

  Type & Solution & $Q_H$ & $Q_V$ \\
  \hline

  H & $\phi=\pm\tanh(2\sqrt{2}x),\ \
\psi_{0}=\pm 2$ &$\pm 1$ & 0 \\
V & $\phi_{0}=\pm1,\ \ \psi=\pm2\tanh(2\sqrt{2}x)$ & 0
& $\pm 1$ \\
D & $\phi=\pm (1/2)\psi,\ \ \psi=\pm\frac{2}{(1+\exp(\pm
4\sqrt{2}x))^{1/2}}$ & $\pm \frac{1}{2}$ &
$\pm \frac{1}{2}$ \\

  \hline
  \hline

\end{tabular}
\end{center}

Since the functions given in Table 1 are not exact solutions of
the coupled equations (\ref{b}) and (\ref{c}), we use them as
initial guesses to find the minimum-energy solutions. Any static
solution has an orbit $f(\phi ,\psi)=0$ in the $(\phi,\psi)$
plane, which begins and ends at one of the true vacua. We
therefore have
\begin{equation}\label{aa}
\frac{df}{dx}=\frac{\partial f}{\partial \phi}\phi'
+\frac{\partial f}{\partial \psi}\psi'=0,
\end{equation}
where prime means differentiation with respect to $x$. The field
equations (\ref{b}-\ref{c}) can be integrated once to yield
\begin{equation}\label{bb}
\frac{1}{2}(\phi ')^2=\int \frac{\partial V}{\partial \phi}d\phi
+C_1,
\end{equation}
and
\begin{equation}\label{cc}
\frac{1}{2}(\psi ')^2=\int \frac{\partial V}{\partial \psi}d\psi
+C_2,
\end{equation}
where $C_1$ and $C_2$ are integration constants. Combining
equations (\ref{aa})-(\ref{cc}), we obtain the following
integro-differential equation for the orbit:
\begin{equation}
\left( \frac{\partial f}{\partial \phi}\right)^2 \left( \int
\frac{\partial V}{\partial \phi}d\phi +C_1\right) =\left(
\frac{\partial f}{\partial \psi}\right)^2 \left( \int
\frac{\partial V}{\partial \psi}d\psi +C_2\right).
\end{equation}

Bazeia et al.\cite{baz} have shown that if the potential can be
written in the form
\begin{equation}
U=\frac{1}{2}\left( F+\phi\frac{\partial F}{\partial
\phi}+\psi\frac{\partial G}{\partial \phi
}\right)^2+\frac{1}{2}\left( G+\phi \frac{\partial F}{\partial
\psi}+\psi \frac{\partial G}{\partial \psi}\right)^2,
\end{equation}
then any solution to the following first order coupled equations
\begin{equation}
\frac{d\phi}{dx}+F+\phi\frac{\partial F}{\partial \phi} +\psi
\frac{\partial G}{\partial \phi}=0,
\end{equation}
and
\begin{equation}
\frac{d\psi}{dx}+G+\psi\frac{\partial G}{\partial \psi} +\phi
\frac{\partial F}{\partial \psi}=0,
\end{equation}
solves the corresponding second order equations. For the coupled
system considered in this paper, we could use this method to
obtain the diagonal (D-type) solutions, provided that
$\phi_0=\psi_0$. In this case, the $F$ and $G$ functions will be
\begin{equation}
F=-\frac{1}{2}\lambda a^2\phi,
\end{equation}
and
\begin{equation}
G=\frac{1}{2}\lambda (\phi^2-a^2)\psi,
\end{equation}
where $a=\phi_0$. For the symmetrical D-type solutions, we have
$\phi^2=\psi^2$, and the exact solution will be
\begin{equation}
\phi^2=\psi^2=\frac{1}{2}a^2\left[ 1+\tanh (\lambda
a^2(x-x_0))\right].
\end{equation}

\begin{figure}[h]
\epsfxsize=10cm \centerline{\epsfbox{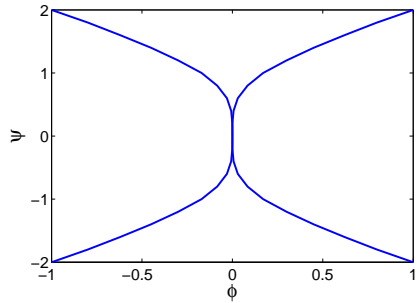}} \caption{The
analytical orbits (\ref{orbiteq}) on the $(\phi , \psi )$ plane.
Compare with the D-type numerical orbits of Figure
5b.\label{orbitfig}}
\end{figure}

{\bf The general D-type orbits are given by\footnote{ We owe these
orbits to the reviewer.}
\begin{equation}\label{orbiteq}
(\phi^2-\phi_0^2) -\phi_0^2\ln \frac{\phi^2}{\phi_0^2}
=(\psi^2-\psi_0^2) -\psi_0^2\ln \frac{\psi^2}{\psi_0^2}.
\end{equation}
These orbits are plotted in Figure \ref{orbitfig} for $\phi_0=1$
and $\psi_0=2$. It can be seen that the D-type numerical orbits of
Figure \ref{5} which minimize the energy functional closely
resemble the analytical orbits of Figure 3.}

In order to find other stable static solutions, we have used the
expressions depicted in Table 1 as the starting "guesses" in a
step-wise variational procedure\cite{1}. In this procedure, the
bi-dimensional spacetime is represented by a grid of spatial and
temporal step-sizes $\epsilon$ and $\delta$, respectively. All the
derivatives in the energy density expression are then written in
their discrete form (e.g. $d\phi/dx\rightarrow
(\phi_i-\phi_{i-1})/\epsilon$). The initial guess is then
repeatedly varied in steps, with the total energy checked at each
step. Only variations which reduce the total energy are accepted
and the rest are rejected. The fields are thus varied toward the
minimum energy solution, were small variations no longer lead to a
decrease in total energy. The minimum energy, static solutions
obtained in this way are plotted in Figure 3. In Figure 4, we have
mapped these solutions into the $(\phi ,\psi )$ plane.

The conservation of $V$ and $H$ type charges (Equation \ref{h})
does not come from a continuous symmetry of the Lagrangian, but
rather it stems from the boundary conditions and topological
considerations \cite{raja}. The system under consideration
respects various types of symmetries: 1) Lorentz symmetry: this
symmetry comes from the invariance of the Lagrangian (\ref{a})
under Lorentz transformations. Since $\phi$ and $\psi$ are scalar
fields, we conclude that if ($\phi (x), \psi(x)$) is a static
solution, then ($\phi(\gamma (x-vt), \psi(\gamma (x-vt)$) where
$v$ is a constant (soliton velocity) and $\gamma =(1-v^2)^{-1/2}$,
is also solution. 2) If ($\phi (x,t), \psi(x,t)$) is a solution,
then ($\pm \phi(x,t), \pm \psi (x,t)$) are solutions, also. 3) For
the special case $\phi_0=\psi_0$, we have dual solutions $\phi
\leftrightarrow \psi$. 4) Parity and time reversal are symmetries
of the system. Therefore if ($\phi(x,t),\psi(x,t)$) is a solution,
then ($\phi(\pm x, \pm t), \psi (\pm x, \pm t)$) are solutions,
too.

\begin{figure}[h]
\epsfxsize=20cm\centerline{\epsfbox{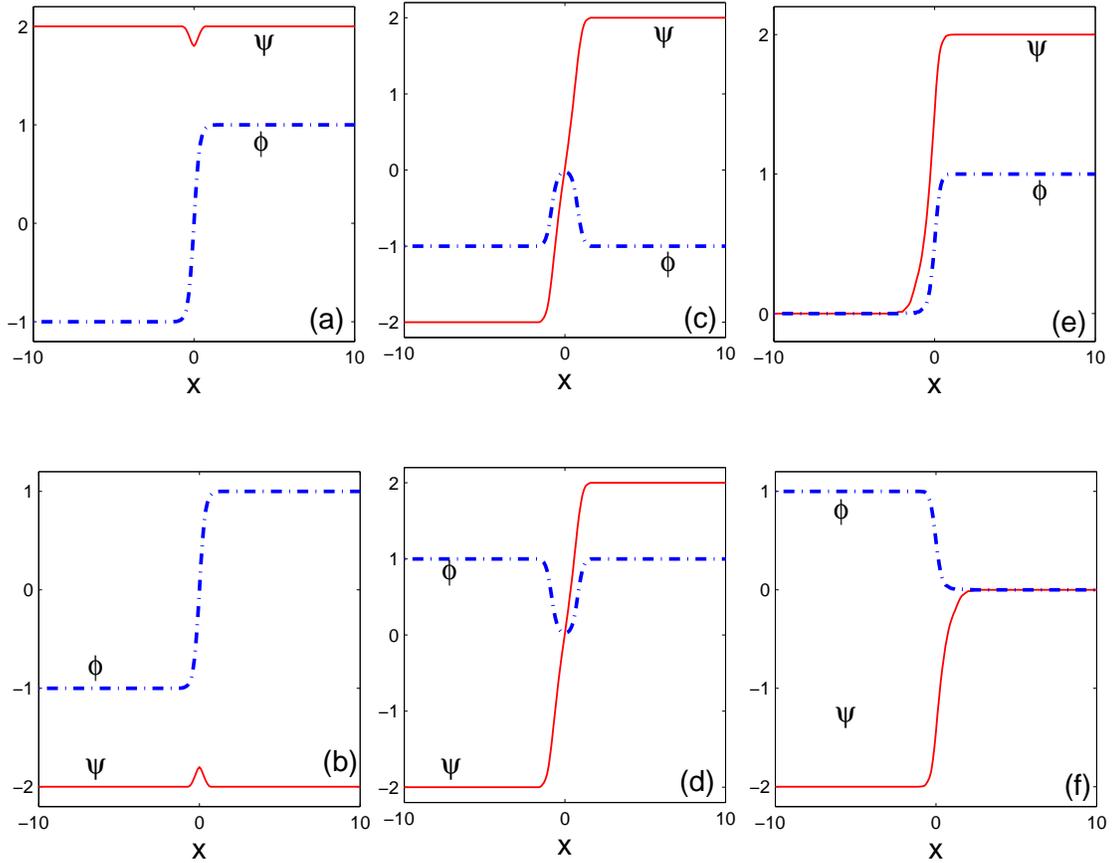}}\caption{Minimum
energy,  static solutions.(a) $H_{BC}$, (b) $H_{DE}$, (c)
$V_{DB}$, (d) $V_{EC}$, (e) $D_{AC}$ and (f) $D_{EA}$. The solid
curves represent $\psi$ and the dash-dotted curves are for
$\psi$.\label{4}}
\end{figure}

\begin{figure}[h]
\epsfxsize=8cm\centerline{\hspace{8cm}\epsfbox{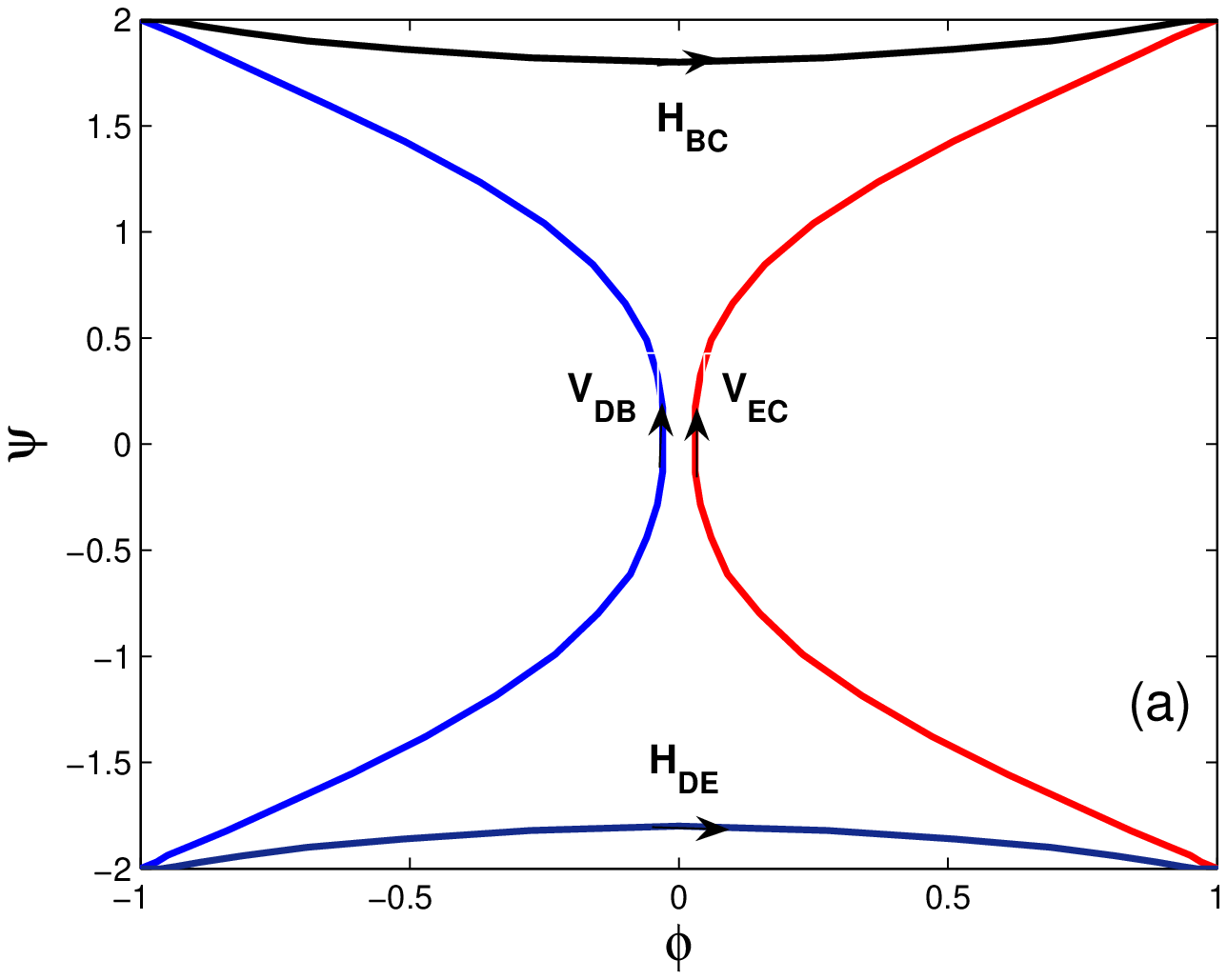}\epsfxsize=8cm\centerline{\hspace{-7cm}\epsfbox{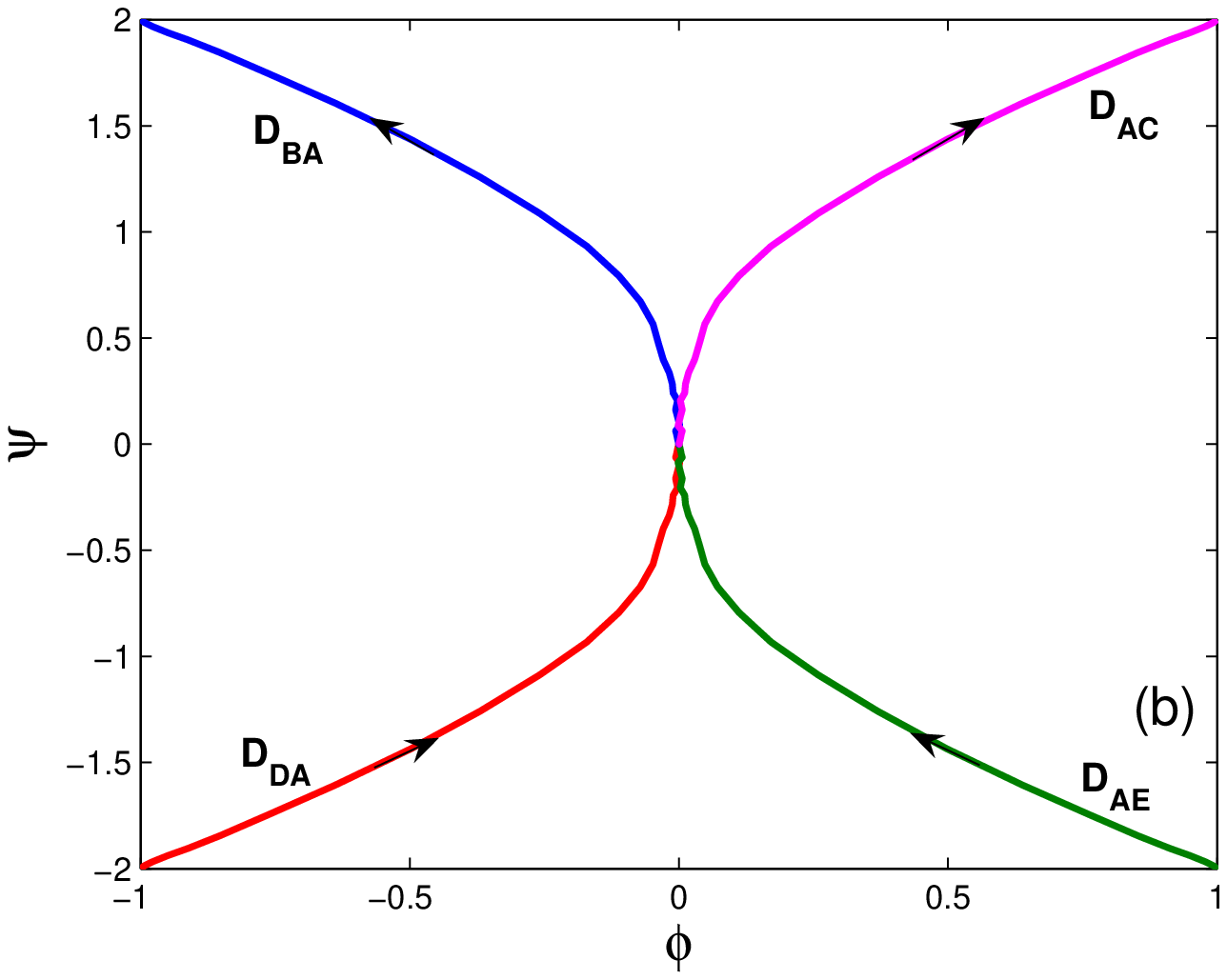}}}
\caption{The minimum energy, static solutions mapped on the
($\phi$,$\psi$) plane: (a) $H_{DE}$, $V_{DB}$, $H_{BC}$ and
$V_{EC}$ and (b) $D_{DA}$, $D_{AC}$, $D_{EA}$ and
$D_{AB}$.\label{5}}
\end{figure}

\section{Soliton Stability and Chiral Decay of $V$-type Solitons}\label{sec4}

Basic properties of the H, V, and D solitons are summarized in
Table 1. It can be seen that all D solitons are degenerate (i.e.
have the same rest energy). This degeneracy results from the
$(\phi , \psi)\leftrightarrow (\pm \phi, \pm \psi)$ symmetry of
the Lagrangian density (2).

\begin{figure}[h]
\epsfxsize=15cm\centerline{\epsfbox{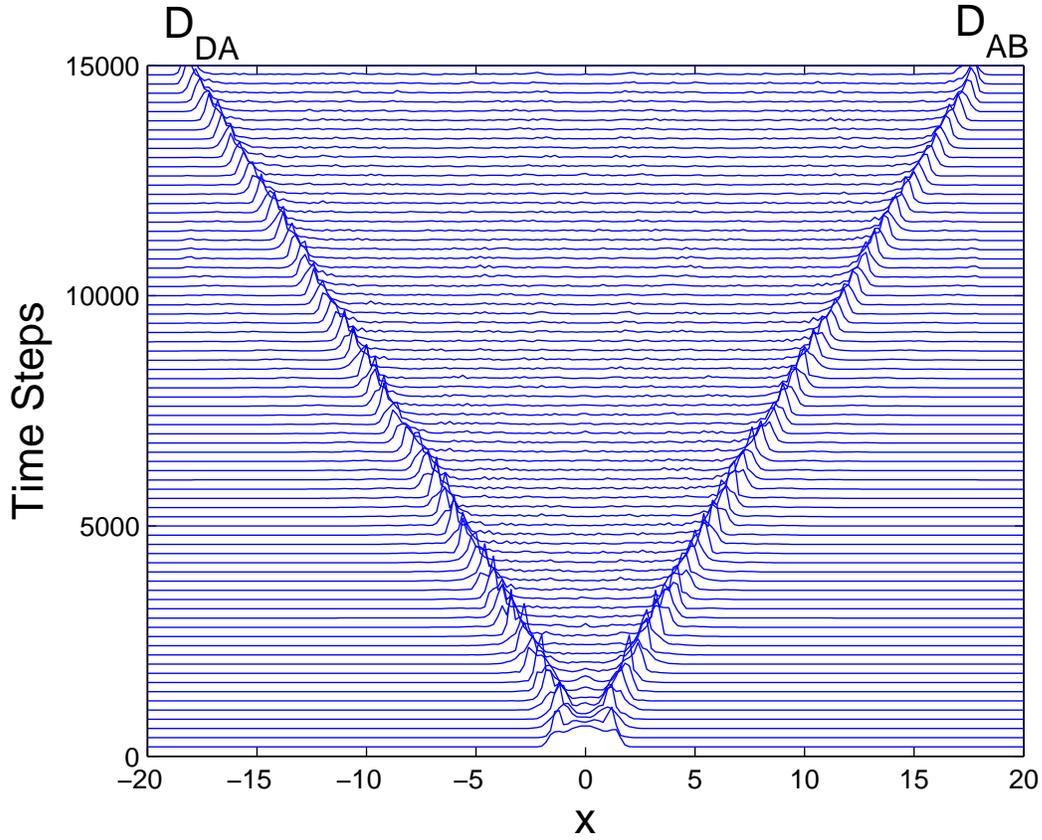}}\caption{Stimulated
decay of a $V_{DB}$ soliton into $D_{DA}+D_{AB}$. Note that the
excess energy is transferred into the kinetic energy of daughter
solitons and also emission of low amplitude waves. Also note that
the decay is chiral, in the sense that  $D_{DA}$ always moves to
the left.\label{decay}}
\end{figure}

Another interesting property of these solitons is that the mass
(rest energy) of the $V_{EC}$ soliton is larger than the sum of
the $D_{EA}$ and $D_{AC}$ solitons, while the topological charge
of $V_{EC}$ is equal to the sum of the $D_{EA}$ and $D_{AC}$
charges. The same is for $V_{DB}$ which is more massive than the
sum of $D_{DA}$ plus $D_{AB}$. The decay of $V_{EC}$ into
$D_{EA}+D_{AC}$ (or $V_{DB}$ into $D_{DA}+D_{AB}$) is therefore
possible with regard to energy and charge conservations. These
decays were not observed to occur spontaneously in our dynamical
simulations, implying that $V_{EC}$ and $V_{DB}$ are classically
metastable configurations. Quantum tunnelling or external
perturbations can -in principle- cause such a decay (see
\cite{1}). In order to show that stimulated decay is in fact
possible in the system under consideration, we have performed the
following simulation: The static vertical soliton $V_{DB}$ is used
as the initial condition in a program which calculates the
dynamics of the system, by solving the coupled PDEs \ref{b} and
\ref{c}. This soliton is excited, by pumping some energy into the
$\phi$ field, via an increase in its amplitude. The system is then
observed to become unstable and decay into $D_{DA}$ and $D_{AB}$.
This simulation is shown in Figure \ref{decay}. Note that the
excess energy is transferred into both the kinetic energy of the
decay products and also emission of some low amplitude waves.

It is interesting to note that the decay of $V$-type solitons is
chiral, in the sense that when $V_{EC}$ decays into $D_{EA}$ and
$D_{AC}$, the former always moves to the left, while the latter
always move to the right. The fact that the reverse never happens
is due to the boundary conditions and topological reasons.
Similarly, for the decay $V_{DB}\rightarrow D_{DA} +D_{AB}$, the
decay product $D_{DA}$ always moves to the left and $D_{AB}$
always moves to the right. This property can be extended to an
arbitrary array of solitons in a multi-soliton system. The
consistency of boundary conditions between successive solitons
permits certain sequences and forbids certain ones. Examples of
allowed and forbidden arrays are the followings:
\[
{\rm Allowed \ \ arrays:}\ \ \ \ \
...D_{DA}D_{AB}H_{BC}V_{CE}H_{ED}...\]
\[
{\rm Forbidden \ \ arrays:}\ \ \ \ \
 ...D_{DA}D_{BA}H_{BC}V_{EC}H_{EC}...
\]

\newpage
\begin{center}
Table 2. General properties of different soliton families. Note
that the soliton masses and stability status depend on the choice
of $\phi_0$ and $\psi_0$. Here, we have assumed $\phi_0=1$ and
$\psi_0=2$.
\begin{tabular}{|c|c|c|c|c|c|}
  \hline
  \hline

  Symbol & Mass & $Q_H$ & $Q_V$ & Stability & Decay Mode\\
  \hline

  $D_{AB}$ &  2.858 & -1/2 & +1/2 & stable & -- \\
  $D_{AC}$ & 2.858 & +1/2 & +1/2 & stable & -- \\
  $D_{DA}$ & 2.858 & +1/2 & +1/2 &  stable & -- \\
  $D_{EA}$ & 2.858 & -1/2 & +1/2 &  stable & -- \\
  $H_{BC}$ & 3.594 & +1 & 0 &  stable & --\\
  $V_{EC}$ & 6.383 & 0 & +1 &  metastable & $V_{EC}\rightarrow D_{EA}+D_{AC}$\\
  $H_{DE}$ & 3.594 & +1 & 0 &  stable & -- \\
  $V_{DB}$ & 6.383 & 0 & +1 &  metastable & $V_{DB}\rightarrow D_{DA}+D_{AB}$\\

  \hline
  \hline

\end{tabular}
\end{center}

\section{concluding remarks}\label{sec5}
We studied a nonlinear system of coupled scalar fields which had
three (H, V, and D) types of stable and metastable soliton
solutions. We started by deriving analytical, static solutions
with definite topological charges. These static solutions,
however, were shown to be unstable and the stable solutions were
obtained numerically, by minimizing the total energy, using a
step-wise variational method. The full, dynamical equations were
used to ensure the stability of these minimum energy solutions.
According to energy and topological charge conservations, the
V-type solutions are expected to decay into D-type ones. This,
however, does not happen spontaneously at the classical level.
Either quantum tunnelling or external triggers are needed for
decays like $V_{EC}\rightarrow D_{EA}+D{AC}$. Conversely, the
formation of $V_{EC}$ from $D_{EA}+D_{AC}$ collision is
endothermic, requiring -at least- the incoming kinetic energy
$\Delta E=6.38 -2\times 2.86=0.66$ in dimensionless units. The
stimulated decay of a $V$-type soliton were calculated
numerically, by pumping some energy into the $\phi$-field. This
decay is exothermic and the excess energy was observed to be
transferred into the kinetic energy of decay products and emission
of low amplitude waves.

 {\bf \large Acknowledgements:} N.R. Acknowledges the support of
 Shiraz University Research Council. Authors would like to thank
 the reviewer for introducing the analytical orbits of Equation
 \ref{orbiteq}.


\begin{thebibliography}{99}
\bibitem{0}G.L. Lamb, Jr., \textit{Elements of Soliton Theory}, John Wiley and Sons, NewYork(1980).
\bibitem{1}N. Riazi, A. Azizi and S. M. Zebarjad, Phys. Rev. {\bf D 66}, 065003 (2002).
\bibitem{2}T. Dauxois and M. Peyrard, \textit{Physics of Solitons}, Cambridge University Press (2006).
\bibitem{3}M. Peyravi,  A. Montakhab, N. Riazi and A. Gharaati, Eur. Phys. J. {\bf B 72}, 269-277 (2009).
\bibitem{4}M. Peyravi, N. Riazi and  A. Montakhab, Eur. Phys. J. {\bf B 76}, 547–555 (2010).
\bibitem{5}N. Riazi, Int. J. Theor. Phys. {\bf 35}, 101 (1996).
\bibitem{6}T.D. Lee, \textit{Particle Physics and Introduction to field Theory}, Harwood, Chur, Switzerland (1981).
\bibitem{dna} L.V. Yakushevich, \textit{Nonlinear Physics of DNA}, Wiley,(2004).
\bibitem{yak} L. V. Yakushevich, A. V. Savin and L. I. Manevitch, Phys. Rev.
{\bf E 66}, 016614 (2002).
\bibitem{cu} S. Cuenda, A. Sanchez, and N.R. Quintero, Physica {\bf D
223}, 214­221 (2006).
\bibitem{pey} M. Peyrard, Nonlinearity {\bf 17}, R1-R40, (2004).
\bibitem{qcd} H. Blas, JHEP, {\bf 0703}, 055, (2007).
\bibitem{ba} D. Bazeia, J. R. S. Nascimento, R. F. Ribeiro, and D. Toledo,
J. Phys. {\bf A 30}, 8157 (1997).
\bibitem{riman} N. Riazi, M. M. Golshan, and K. Mansuri, Int. J. Theor. Phys. Group Theor. Non.
Opt. {\bf 7}, 91 ((2001).
\bibitem{7}L.H. Ryder, \textit{Quantum Field Theory}, Cambridge University Press (1985).
\bibitem{8}M. Guidray, \textit{Gauge Field Theories, An Introduction with Application}, John Wiley and Sons, NewYork(1991).
\bibitem{10}R. H. Goodman and R. Haberman, Siam J. Applied Dynamical Systems
{\bf  4}, No. 4, 1195, (2005).
\bibitem{11}S. Hoseinmardy and N. Riazi, IJMPA. {\bf 25},  3261, (2010).
\bibitem{baz} D. Bazeia, M.J. dos Santos, and R.F. Ribeiro,
Phys. Lett. A, {\bf 208}, 84, (1995).
\bibitem{raja} R. Rajaraman, Soltions and Instantons, Elsevier,
B.V. (1989).
\end{thebibliography}
\end{document}